\documentclass[a4paper,10pt,twoside]{cpc-hepnp}

\usepackage{graphicx}
\usepackage{multicol}
\usepackage{amssymb,bm,mathrsfs,bbm,amscd}
\usepackage[tbtags]{amsmath}
\usepackage{lastpage}

\begin{document}
\title{Oxygen quenching in LAB based liquid scintillator and nitrogen
bubbling model
\thanks{Supported by Nation Natural Science Foundation of China (211202037)}
}


\author{%
XIAO Hua-Lin  $^{1}$\email{xiaohl@mail.bnu.edu.cn}
 \quad DENG Jin-Shan
}

\maketitle




\address{%
College of Nuclear Science and Technology, Beijing Normal
University, Beijing 100875, PR China }

\begin{abstract}
The oxygen quenching effect in Linear Alkl Benzne (LAB)  based
liquid scintillator (LAB as the solvent, 3 g/L 2, 5 diphe-nyloxazole
(PPO) as the fluor and 15 mg/L $p$-bis-($o$-methylstyryl)-benzene
(bis-MSB) as the $\lambda$-shifter) is studied by measuring the
light yield as the function of the nitrogen bubbling time. It is
shown that the light yield of the fully purged liquid scintillator
is increased by 11\% at the room temperature and the room atmosphere
pressure. A simple nitrogen bubbling model is proposed to describe
the relationship between the relative light yield (oxygen quenching
factor) and the bubbling time.
\end{abstract}



\footnotetext[0]{\hspace*{-2em}\small\centerline{\thepage\ --- \pageref{LastPage}}}%

\begin{multicols}{2}


\section{Introduction}


It has been shown experimentally and theoretically that the presence
of oxygen in the liquid scintillator (LS) can lower the light yield,
modify the fluorescence pulse shape, shorten the attenuation length
and decrease the positron annihilation
lifetime\cite{pulseshape,attenuationlength,positronlifetime}. For
most aromatic molecules, the quenching of the electronically singlet
state (S$_{1}$) leads to the formation of triplet state (T$_{1}$).
The oxygen molecule is somewhat special. Its ground state is a
triplet and the the next state is a singlet lying about 0.98 eV over
the ground state\cite{stateenergy}; Oxygen molecules in aromatic
molecules can absorb the energy of singlet state of aromatic
molecules, and make the spin allowed transition to the triplet
state. This decreases the fluorescence.
Such a transition only occurs in aromatic molecules  which have a
energy gap of S$_{1}$ --T$_{1}$ greater than 0.98 eV. Most of
aromatic molecules meet this requirement. In most experiments, the
dissolution of oxygen into LS is undesirable. Since this brings
uncertainties to the experiments. Hence, the oxygen quenching effect
should be well studied and determined.


Usually, there are three ways to eliminate the dissolved oxygen from
solutions: 1) the vacuum distillation; 2) the
ultrasonic\cite{ultrasonic}; 3) the nitrogen(or argon)
bubbling\cite{pulseshape}. In neutrino experiments, large quantity
of liquid scintillator is required. The most  economical and
practicable way to eliminate oxygen in LS is the nitrogen (or argon)
bubbling.

Linear alkyl benzne  (LAB), which is composed of a linear alkyl
chain of 10-13 carbon atoms attached to a benzene ring
 ,
is a low cost product of petrochemical industry and is often used as
the material of detergent.
 Its aromatic structure makes it be useful as a scintillator
solvents inherently. It has many appealing properties, including the
high flash point ($130\,^{\circ}\mathrm{C}$), the low toxicity, the
high light yield and excellent transparency\cite{liujinchang}. LAB
based liquid scintillator will serve as the antineutrino target in
the Daya Bay neutrino experiment\cite{dingyayun}.

In this work, we measured the effect of oxygen on the light yield of
LAB LS, and built a nitrogen bubbling model to describe the
relationship between relative light yield (the oxygen quenching
factor) and the nitrogen bubbling time. Parameters in the model were
determined by our experimental data.

\section{Nitrogen bubbling model}
    When LS is exposed to the air, oxygen molecules dissolved in the LS  exchanges with those in the air.
This process is in dynamical equilibrium. It is reasonable to assume
that the oxygen dissolved
 in the un-bubbled LS is saturated due to its long time contact with the air.
This means that the number of oxygen molecules dissolving into the
LS is equal to those escaping from the LS. When LS is flushed with
nitrogen, the oxygen partial pressure in the nitrogen bubble, which
presents in LS, can be thought to be zero. Therefore, oxygen
molecules will escape from LS and enter the nitrogen bubbles. The
dissolution of oxygen molecules into the LS can be ignored in the
interface of nitrogen and LS. Since that the oxygen molecules
diffusion rate into LS is much higher than the oxygen escaping rate,
which denotes that oxygen molecules are uniformly distributed in LS;
It is reasonable to assume that the oxygen escaping rate is
proportional to the contact area of bubbles with LS and the oxygen
partial pressure in LS which is proportional to the oxygen
concentration. Then, the equation describing the variation of the
oxygen molecule number
in LS, $dN/dt$, is given by\\
\begin{equation}
\label{eq:dn1} \frac{dN}{dt}=-k_{e}[Q]S,
\end{equation}
where $[Q]$ is the oxygen concentration dissolved in LS; $k_{e}$ and
$S$ are the oxygen escaping rate and nitrogen--LS contact area,
respectively. The oxygen concentration can be expressed as the
following:
\begin{equation}
\label{eq:oxygencontration} [Q]=\frac{N}{V_{s}},
\end{equation}
where  $N$ and $V_{s}$ are the oxygen molecule number in LS and the
LS volume, respectively. Hence, Eq.~(\ref{eq:dn1}) can be rewritten
as
\begin{equation}
\label{eq:dn2} \frac{dN}{dt}=- \frac{k_{e} N}{V_{s}} S.
\end{equation}

Then, the variation of the oxygen molecule number  in LS is
 \begin{equation}
 \label{eq:dn}
 \frac{dN}{N}=- \frac{k_{e} S}{V_{s}} dt.
 \end{equation}
Eq.~(\ref{eq:dn}) can be written in the integration form:
 \begin{equation}
 \label{eq:N}
N=N_{0}\exp \Big(-\frac{k_{e}  S }{V_{s} } t \Big ),
\end{equation}
where $N_{0}$ is the oxygen molecule number in LS without bubbling.

At low concentration, the quenching of fluorescence by a quencher in
solution can be described by the well-known Stern-Volmer
relationship,
\begin{equation}\label{eq:sternvolmer}
\frac{I_{0}}{I}=1+k_{Q} [Q],
\end{equation}
where $I_{0}$ is the intensity or rate of fluorescence without a
quencher present, $I$ is the intensity or the rate of fluorescence
with a quencher, $[Q]$ is the quencher concentration dissolved in
LS, and $k_{Q}$ is quenching constant.


Hence, the relative light yield as the function of the bubbling time
is given by
\begin{equation}
\label{eq:f1} \frac{I_{0}}{I}=1+\frac{k_{Q} N_{0}}{V_{s}}\exp \Big
(-\frac{k_{e}  S }{V_{s}} t \Big ).
\end{equation}
Note that $N_{0}/V_{s}$ is the saturated concentration of oxygen
which is dependent of temperature and atmospheric pressure; At fixed
temperature and atmospheric pressure, $N_{0}/V_{s}$ is a constant.
Denote the $k_{Q} N_{0}/ V_{s}$ item by a constant $A$. Then,
Eq.~(\ref{eq:f1}) turns to be
\begin{equation}
\label{eq:f11} \frac{I_{0}}{I}=1+A \exp \Big (-\frac{k_{e} S }{V_{s}
} t \Big ).
\end{equation}

The oxygen quenching factor, $f_Q$, is defined as the light yield of
LS with oxygen to that without oxygen, i.e. $I/I_0$.
Eq.~(\ref{eq:f11}) can be expressed in the form of $f_Q$,

\begin{equation}
\label{eq:f12} f_Q=1/(1+A \exp  (-\frac{k_{e} S }{V_{s} } t  )).
\end{equation}
\begin{center}
  \includegraphics[width=0.15\textwidth]{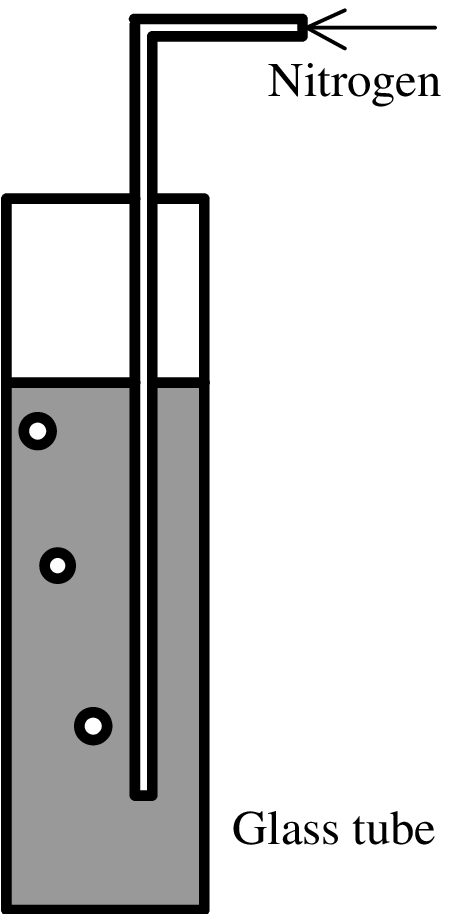}\\
  \figcaption{\label{fig:bubblesetup}Bubbling setup}
\end{center}

The bubbling setup can be illustrated by the simplified plot in Fig.
\ref{fig:bubblesetup}.  The nitrogen--LS contact area consists of
two parts: the area of the nitrogen bubble surface and the contact
surface area of liquid level, i.e.
 \begin{equation}
 \label{eq:dn3}
 S=n S_{b}+S_{l},
 \end{equation}
where $n$ is the average number of nitrogen bubble present in the
LS. $S_{b}$ and $S_{l}$ are the average surface area of bubbles and
the area of the liquid level, respectively. Then, Eq.~(\ref{eq:f12})
can be rewritten as
\begin{equation}
\label{eq:f121} f_Q=1/(1+A \exp  (-k_{e}( n S_{b}+S_{l}) t/V_{s})).
\end{equation}

The parameters $A$ and $k_{e}$ are constants  depending on the
temperature and pressure for specific LS. In the following sections
we will evaluate $A$ and $k_{e}$ for LAB LS at the room temperature
and the atmospheric pressure experimentally.

\section{Experiment}

 In order to observe the light output variation due to the oxygen
quenching, sets of  samples of LS (LAB as the solvent, 3 g/L PPO as
the fluor and 15 mg/L bis-MSB as the wavelength shifter) were
bubbled with different time. Fig.~\ref{fig:bubblesetup} shows the
bubbling setup. Six sets of LS samples (50 ml for each sample) were
 used. One set of samples was not bubbled,  and the other five
were bubbled with high purity nitrogen for 5 min (200 ml), 12.5 min
(500 ml), 18.7 min (750 ml), 25 min (1000 ml)  and 31.25 min (1250
ml), respectively (the numbers in the brackets are the nitrogen
volumes). Nitrogen flow rate was precisely controlled at 40 ml/min
by a flowmeter. The bubbles present in LS can be thought to be
spherical. The bubble number appears in the tube can be easily
counted and the diameters of bubbles can be measured with rulers. In
our condition, 4 bubbles, with 4 mm in diameter, present in LS. The
diameter of the tube is 23 mm. Hence, the LS--nitrogen contact area
is 616.5 mm$^2$.
 \begin{center}
  \includegraphics[width=0.4\textwidth]{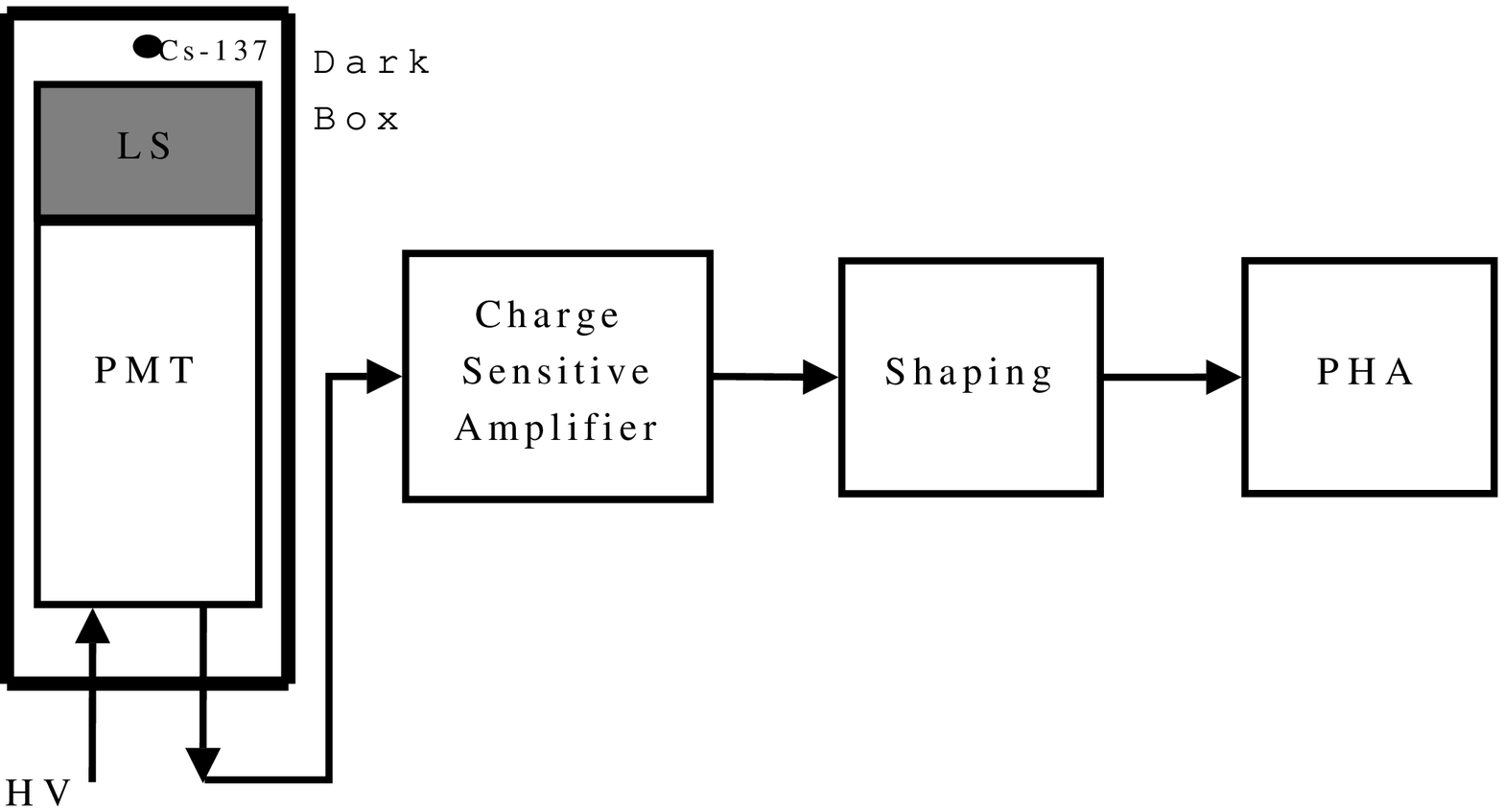}\\
  \figcaption{\label{fig:setup}Schematic view of the experimental setup.}
\end{center}

The setup shown in Fig.~\ref{fig:setup} was used to measure the
light output. LS  was encapsulated in a cylindrical teflon cell (5
cm in diameter and 2.5 cm in height). The
end of the cell was terminated with UV glass 
which was coupled to a 2-inch high energy resolution  PMT (Hamamatsu
CR105). The cell and PMT were placed in a dark steel box. The cell
was exposed to a $^{137}$Cs $\gamma$ -ray source.

 \begin{center}
  \includegraphics[width=0.4\textwidth]{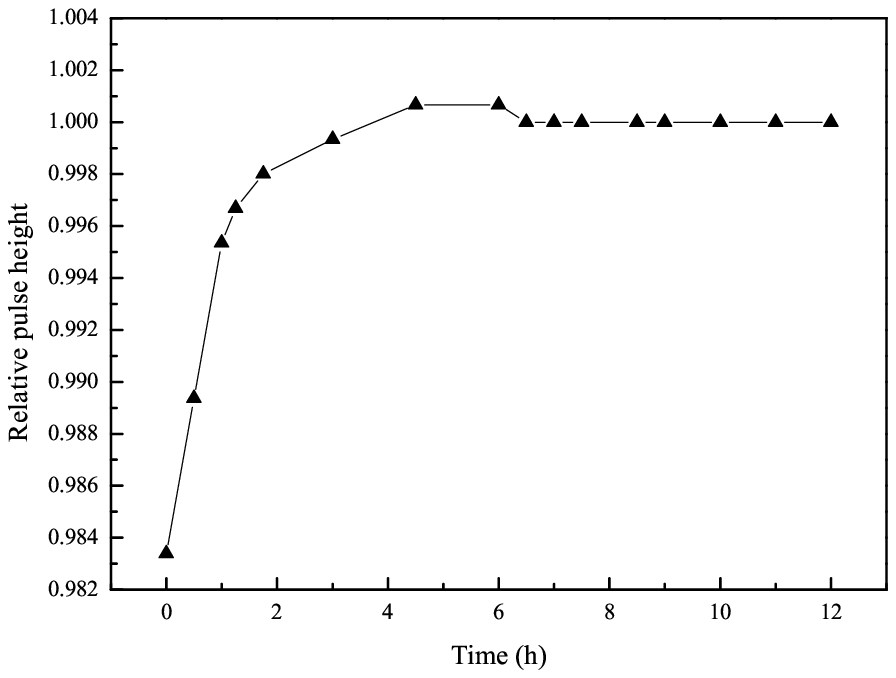}\\
  \figcaption{\label{fig:stability}System stability in 12 hours.The stability was tested by a LED driven by a pulse generator}
\end{center}

The charge of PMT output pulse  was firstly coversed to amplitude by
a charge sensitive amplifier and then shaped by shaping filter.
Finally, the shaped signal amplitude was analyzed by a pulse height
analyzer. System stability has been measured by means of a LED
driven by pulser generator. Fig.~\ref{fig:stability} shows the
system stability. System trended stable after 6 hours burning.
System was burned about 12 hours before data acquisition in our
experiment.


\section{Data Analysis and Results}
\begin{center}
  \includegraphics[scale=0.7]{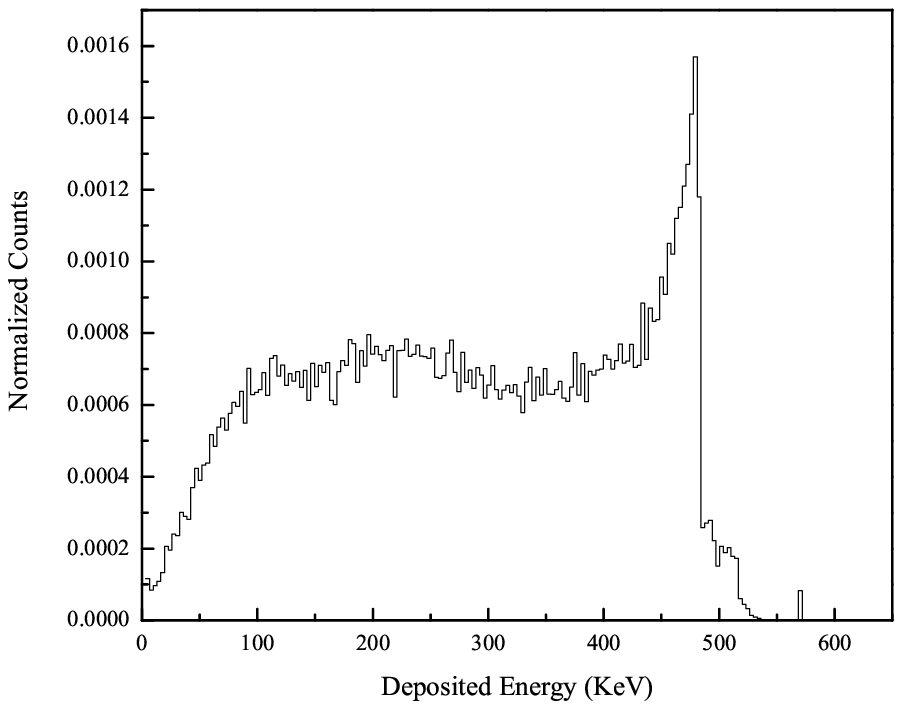}\
  \figcaption{\label{fig:gresp}Energy response of LS to the
  $^{137}$Cs $\gamma$-ray source simulated by GRESP.
The counts are normalized to the number of $\gamma$-rays emitted
from the $^{137}$Cs
  source. }
\end{center}

Energy spectrum $N(E)$ of the Compton scattering electrons is
generated by means of Monte Carlo GRESP code. Fig.~\ref{fig:gresp}
shows the simulation result. It should be noted that the resolution
smearing is not considered in the GRESP Monte Carlo code. We
considered the resolution smearing similar to the way described in
Ref \citep{smearing}. The ``realistic '' Monte Carlo spectrum can be
obtained from the convolution of the simulation spectrum with the
system response function,
\begin{equation}\label{eq:convolution}
N ^{MC}(H)=\int R(H,L) N_L(L)dL,
\end{equation}
where $R(H,L)$ is the response function, $H$ is the ADC channel and
$N_L(L)$ is the spectrum of light output. For electrons (above about
50 KeV), the light output $L$, the energy emitted as fluorescence,
is proportional to the energy $E$ deposited in the LS\cite{jbbirks}.
 , i.e.
\begin{equation}\label{eq:lequalssE}
L=SE,
\end{equation}
where $S$ is the absolute scintillation efficiency. The light output
$L$ can be defined to be in units such that it is equal to the
electron energy $E$\cite{lse}, i.e $S=1$ and
\begin{equation}
L(E)=E.
\end{equation}
When oxygen presents in LS, the light yield will decrease. Let the
quenching factor, i.e. the light yield with the oxygen quenching to
that without the oxygen quenching, $I/I_{0}$, be $f_{Q}$. Then the
light output can be rewritten as
\begin{equation}\label{eq:lfqse}
 L=f_{Q}E.
\end{equation}
It should be reminded that $f_Q$ is 1 for LS without a quencher.
Considering that the spectrum of deposited energy is $N(E)$, the
spectrum $L$, $N_L(L)$, is $N(L/ f_{Q}S)/f_{Q}$. Assuming that the
response $R(H,L)$ of the detector for the fixed light output is
Gaussian,
\begin{equation}
\label{eq:guassian} R(H,L)=B \exp \Big(- \frac{(H-cL)^{2}}{2
\sigma_{cL}^{2}} \Big),
\end{equation}
where $c$ is the light output to ADC channel conversion factor,  and
$B$ is the normalization factor between the ``realistic''  Monte
Carlo spectrum and the experimental spectrum. $\sigma _{cL}$ can be
expressed in the form of the system resolution,
\begin{equation}\label{eq:sigmacl}
\sigma _{cL}=\frac{cL }{2\sqrt{2\ln2}} \rho,
\end{equation}
where $\rho=\Delta(cL)/cL $ is the detector resolution, i.e. FWHM.
and the background can be thought to be
expositional\cite{liujinchang},
\begin{equation}\label{eq:background}
N ^{BG}(H)=c_{1}\exp\Big(-c _{2} H +c_{3}\Big),
\end{equation}
where $c_{1}$, $c_{2}$, $c_{3}$ are parameters of the background.
Then, the expected experimental spectrum can be written as
\begin{eqnarray}
\label{eq:realisticmc} N ^{MC}(H)=B \int
\exp \Big(-\frac{(H-cf_{Q}E)}{2\sigma _{E} ^{2}}\Big )N(E)dE
\nonumber\\
+c_{1}\exp \Big(c_{2}H+c_{3}\Big),
\end{eqnarray}
with
\begin{equation}
\sigma_{E}=\frac{cf_{Q}E \rho}{2\sqrt{2 \ln 2}}.
\end{equation}

\begin{center}
  \includegraphics[width=.4\textwidth]{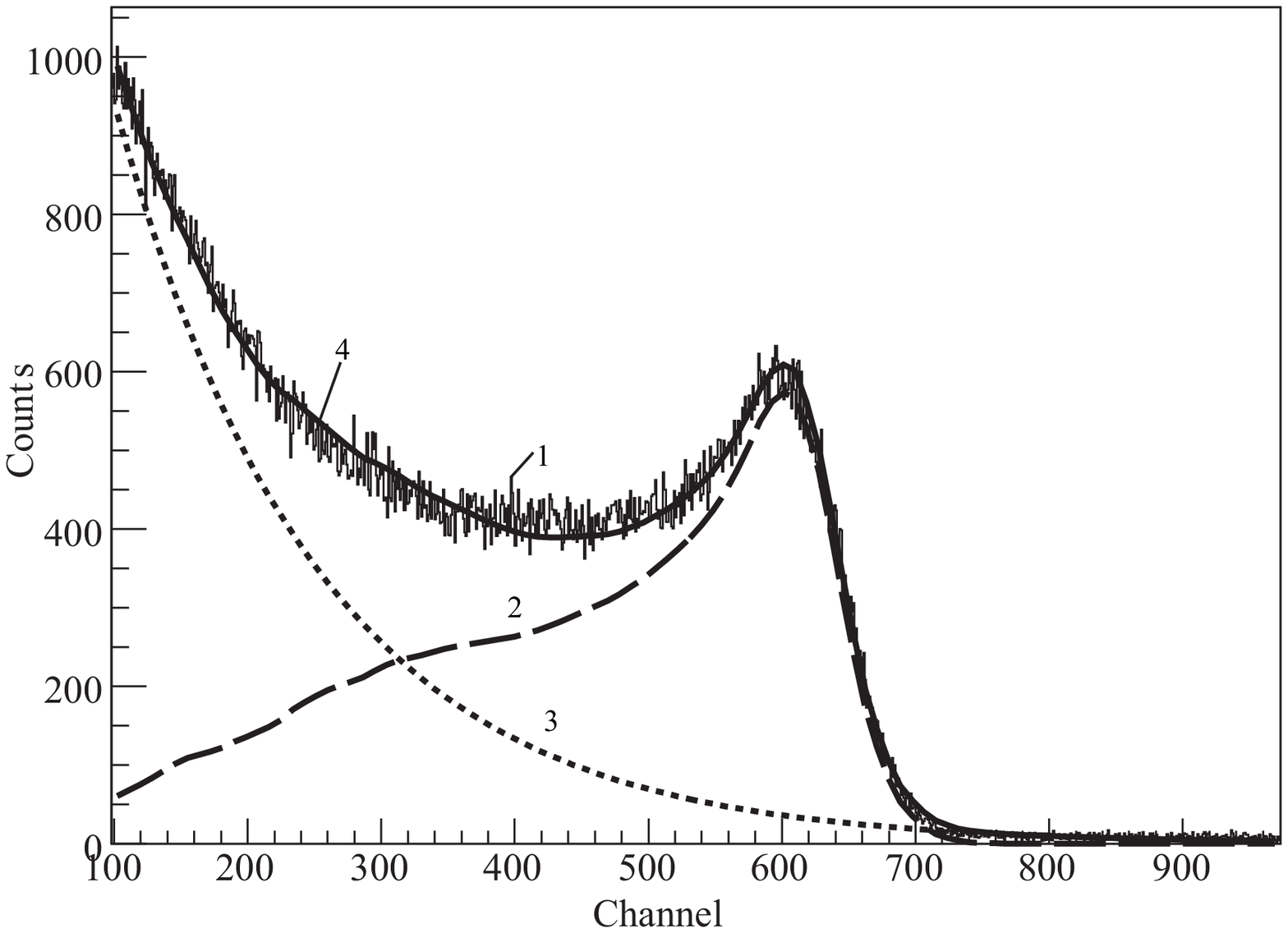}\\
  \figcaption{\label{fig:GrespFit}
    (1) Experimental spectrum, (2) ``realistic" Monte Carlo spectrum
        ,(3) Background defined in Eq.~(\ref{eq:background}), (4) expected experimental  spectrum defined in
    Eq.~(\ref{eq:realisticmc})  for $^{137}$Cs.
   }
\end{center}
 The expected experimental spectrum is determined by $B, c_{1},  c_{2},
c_{3}$,  $c$, $f_{Q}$ and the detector resolution $\rho$. $c$
depends on the optical properties of the cell, PMT and electronics.
It is independent of quenching and can be treated as the a constant
in our experiment. Hence the item $f_{Q} c$, which is proportional
to the quenching factor, can be taken as one free parameter. To
evaluate the free parameters we used the ROOT package (CERN data
analysis package) to fit the experimental spectrum with
Eq.~(\ref{eq:realisticmc}). Fig.~\ref{fig:GrespFit} shows the fit
result of the background, the realistic spectrum, and the expected
spectrum for the experimental spectrum. The items $f_{Q}c$  for
different LS samples can be obtained by fitting their experimental
spectrums with Eq.~(\ref{eq:realisticmc}). With the increase of the
nitrogen bubbling time, the light output  increased. We assume that
oxygen was fully bubbled, namely no oxygen quenching effect remains
in LS, when the light yield change little. It should be reminded
that the quenching factor, $I/I_0$, for full bubbled LS is 1. $c$
was established from the fit result of the full bubbled LS. Then,
the value of $f_Q$ for different LS samples were determined.
Fig.~\ref{fig:bubbletime} shows the $f_{Q}$ for six LS samples. The
solid line in Fig.~\ref{fig:bubbletime} shows the fits with
Eq.(\ref{eq:f12}). From the experimental result, we know that the
light yield  is increased by about 11\% (20$^\circ \mathrm{C}$) by
means of removing the oxygen in LS.
\begin{center}
  \includegraphics[width=.4\textwidth]{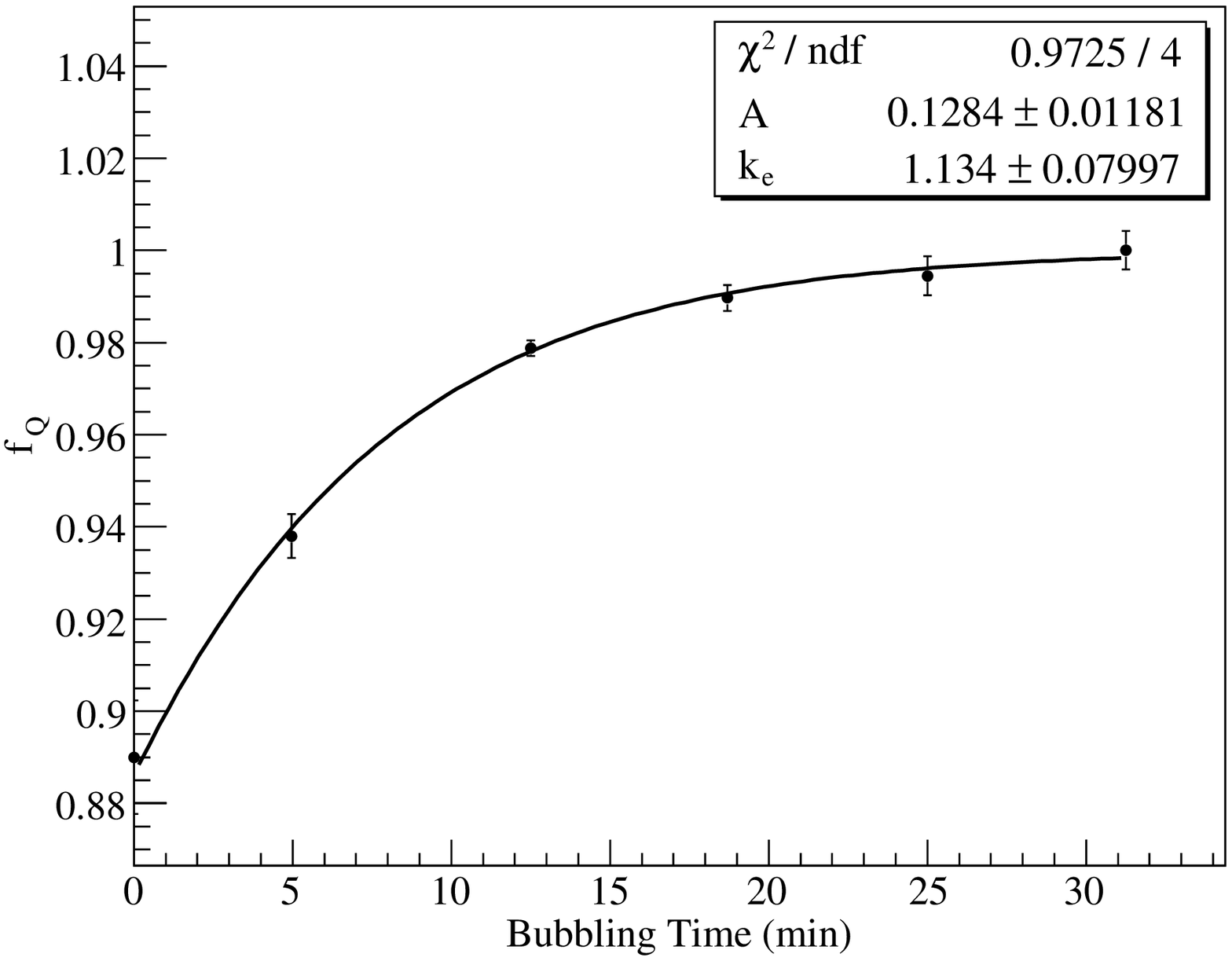}\\
  \figcaption{\label{fig:bubbletime}$f_Q$ as a function of the bubbling time. The
 solid curve shows the fits with Eq.~(\ref{eq:f12}).}
\end{center}

\section{Conclusions}
 The oxygen quenching in LAB liquid scintillator and the degassing
model have been studied. From the experiment, we know that LAB LS
light yield is increased by 11\% by fully removing oxygen at
20$^\circ \mathrm{C}$. Moreover, we proposed a model to determined
the relationship between the light yield and the bubbling time in
this paper. The parameters in the model have been fixed
experimentally.

\section{Acknowledgements}
This work is supported by the Natural Science Foundation of China
(211202037). The authors would like to thank Xin-Heng Guo, Jun Cao,
Bing-Lin Yong and Xi-Chao Ruan and  people from Institute of High
Energy Physics who gave help and suggestion for our work.
\end{multicols}
\vspace{-2mm} \centerline{\rule{80mm}{0.1pt}} \vspace{2mm}
\begin{multicols}{2}

\end{multicols}

\clearpage

\end{document}